\documentstyle[twocolumn,tighten,aps,epsfig]{revtex}
\begin{document}
\draft

\def\Etrans{E_{\rm trans}}
\def\coal{$A+A\to A$}
\def\annil{$A+A\to0$}
\def\av#1{\langle#1\rangle}
\def\erfc{{\rm erfc}}
\def\gsim{\mathrel{\raise.36ex\hbox{$>$\kern-.8em\lower1ex\hbox{$\sim$}}}}
\def\lsim{\mathrel{\raise.36ex\hbox{$<$\kern-.8em\lower1ex\hbox{$\sim$}}}}
\def\ds{{d_{\rm s}}}
\def\dw{{d_{\rm w}}}
 
\title{Diffusion-Limited Coalescence and Annihilation in Random Media}
\author{Catalin Mandache and Daniel ben-Avraham\footnote
  {{\bf e-mail:} qd00@clarkson.edu}}

\address{Physics Department, and Clarkson Institute for Statistical
Physics (CISP), \\ Clarkson University, Potsdam, NY 13699-5820}
\maketitle

\begin{abstract} 
We study the kinetics of diffusion-limited coalescence, \coal, and
annihilation, \annil, in random media consisting of disconnected domains of
reaction.  Examples include excitons fusion and annihilation in porous matrices
and along polymer chains.  We begin with an exact analysis of $A+A\to A$ in a
finite segment.  This result is applied to coalescence in a random
distribution of segment lengths, and the implications for coalescence and
annihilation in percolation clusters and other confined geometries are then
derived by means of scaling techniques.
\end{abstract}


\section{Introduction}
Diffusion-limited reactions are far less
understood than their reaction-limited counterpart.  Simple
reaction models such as diffusion-limited coalescence,
\coal, and annihilation, \annil, have attracted much recent attention 
\cite{Krebs,Henkel,Simon,Droz,Bramson,Torney,Spouge,Peliti,Racz,Lushnikov,%
Takayasu,Privman,benavraham90,Albuquerque},
since their kinetics in one-dimensional space may be analyzed exactly, and
they shed light on other less tractable reaction schemes.  Moreover,
coalescence and annihilation serve as models of exciton dynamics in several
real systems.  Examples include photoexcited solitons and polarons in MX
chain compounds, such as  [Pt(en)$_2$][Pt(en)$_2$Cl$_2$](BF$_4$)$_4$ ((en)
denotes ethylenedyamine)
\cite{kuroda97,okamoto97}, and fusion of photogenerated excitons in
tetramethylammonium manganese trichloride (TMMC) chains \cite{kroon97}. 
Kroon and Sprik \cite{kroon97} review many other systems which are aptly
modeled by  diffusion-limited coalescence or annihilation in one dimension.

Most theoretical studies to date assume an infinite medium, however, rich
kinetic behavior arises from the finite size of the substrate in question. 
For times that exceed the typical diffusion time needed to span the size of the
sample, finite-size effects set in and dramatically affect the kinetics.  We
argue that when the system consists of a collection of reaction-domains of
varying sizes, this might lead to a stretched exponential decay in the
concentration of particles, as opposed to the usual power-law decay.  For
example, disorder in a system of polymers might effectively fracture the
medium into separate reaction zones.  Indeed, stretched exponential decay of
photogenerated excitons in MX compound chains \cite{kuroda97,okamoto97} and in
TMMC chains \cite{kroon97} is a well documented fact.

The problem generalizes to systems of higher dimensionality.  Triplet exciton
annihilation in naphtalene-h$_8$ embedded in porous matrices of
napthalene-d$_8$ has been studied by Kopelman \cite{Kopelman}.  Here the
geometry of the reaction zones and their characteristic sizes might be well
described by percolation theory.  Kopelman \cite{Kopelman} has studied
exciton dynamics in several systems, spanning an impressive range of different
geometries.

In this paper we deal with diffusion-limited coalescence, or annihilation, in
random media --- media consisting of disconnected reaction domains with a given
distribution of sizes.  We begin with an exact derivation of the kinetics of
coalescence, \coal, in a finite segment of length $L$, in section~II.  At
early times, $t\ll1/Dc_0^2$ ($D$ is the diffusion coefficient of the
particles, and $c_0$ is their initial concentration), few reactions take place
and the concentration is nearly constant; at the intermediate time regime of
$1/Dc_0^2\ll t\ll L^2/D$ the concentration decays in power-law fashion,
$c\sim1/\sqrt{Dt}$; and in late times, $L^2/D\ll t$, the decay is exponential,
$c\sim\exp[-(2\pi^2D/L^2)t]$.  In section~III, we use these results to derive
the kinetics of diffusion-limited coalescence of particles confined to a set of
segments characterized by the distribution of their lengths, $\psi(L)$, as
might be the case for exciton dynamics in polymer compounds.  For a completely
random distribution, $\psi(L)=\rho^{-1}\exp(-\rho L)$, we find that the long
time asymptotic decay is stretched-exponential, $c\sim\exp(-t^{1/3})$, similar
to the long time behavior in the famous problem of trapping \cite{Donsker}. 
The earlier time regime is dominated by a somewhat faster
stretched-exponential decay,
$c\sim\exp(-t^{1/2})$.  Coalescence and annihilation in percolation clusters
is addressed in section~IV.  Here the situation is more complicated, for the
reaction domains are fractal and diffusion is then anomalous
\cite{anomalous}.  Our analysis consists of scaling arguments, motivated by
the more firmly grounded results in one dimension.  We conclude with a
summary and discussion, in section~V.  

\section{Coalescence in a segment} 

Consider the diffusion-limited coalescence process, \coal, where the particles
diffuse with diffusion coefficient $D$ and merge instantaneously upon
encounter, taking place in a segment of length $L$.
An exact analysis is possible through the method of Empty Intervals, known also
as the method of Inter-Particle Distribution Functions (IPDF)
\cite{benavraham90}.  The key parameter is $E(x,t)$ --- the probability that an
arbitrary interval of length $x$ is empty (contains no particles), at time
$t$.  $E(x,t)$ satisfies the diffusion equation \cite{benavraham90}:
\begin{equation}
\label{dE/dt}
{\partial\over\partial t}E(x,t)=2D{\partial^2\over\partial x^2}E(x,t)\;.
\end{equation}
The coalescence reaction imposes the boundary condition
\begin{equation}
\label{E=1}
E(0,t)=1\;,
\end{equation}
and, because $E$ is a {\em probability}, also
\begin{equation}
\label{E>0}
E(x,t)\geq 0\;.
\end{equation}
Finally, if there are any particles in the segment then
\begin{equation}
\label{E(L)=0}
E(L,t)=0\;. 
\end{equation}

{}From $E(x,t)$ one can compute various useful quantities.  For example, the
particle concentration is
\begin{equation}
\label{conc}
c(t)=-{\partial E(x,t)\over\partial x}|_{x=0}\;,
\end{equation}
and the conditional probability that the next nearest particle to a given
particle is at distance $x$ (the IPDF), is
\begin{equation}
p(x,t)=c(t)^{-1}{\partial^2\over\partial x^2}E(x,t)\;.
\end{equation}

The steady state solution of eq.~(\ref{dE/dt}) ({\it i.e.}, $\partial
E/\partial t=0$), with the boundary conditions (\ref{E=1}), (\ref{E>0}), and
(\ref{E(L)=0}), is
\begin{equation}
\label{Es}
E_s(x)=1-{x\over L}
\end{equation}
Clearly, if there is initially at least one particle present, a single particle
would remain at the end of the process, unable to undergo any further
reactions.  This is confirmed by eq.~(\ref{Es}), and using~(\ref{conc}):
\begin{equation}
c_s={1\over L}\;.
\end{equation}
In systems consisting  of various reaction domains, this gives
rise to a residual concentration at the end of the process (Appendix~A).  Our
goal is to describe the approach of the concentration to the residual limit.
 
For the full time analysis, we rewrite the empty interval probability as a sum
\begin{equation}
E(x,t)=E_s(x)+\Etrans(x,t)\;,
\end{equation}
where $\Etrans$ symbolizes the transient behavior of interest.  $\Etrans$
satisfies the same equation as $E$, eq.~(\ref{dE/dt}), but the boundary
condition~(\ref{E=1}) is replaced by
\begin{equation}
\label{bc:Etrans}
\Etrans(0,t)=0\;.
\end{equation}
A spectral decomposition then yields
\begin{equation}
\label{Etrans}
\Etrans=\sum_{n=1}^{\infty}a_n\sin({n\pi x\over L})e^{-\lambda_nt}\;;\quad
\lambda_n={2n^2\pi^2D\over L^2}\;,
\end{equation}
where the $\{a_n\}$ are determined from initial conditions.

For practical purposes, we consider the case where the segment
is initially filled with particles at concentration $c_0$.  We assume that
the particles are randomly (Poisson) distributed, and hence the initial empty
interval probability, given that the segment contains at least one
particle, is
\begin{equation}
E(x,0)={(e^{-c_0x}-e^{-c_0L})\over1-e^{-c_0L}}\;,
\end{equation}
and
\begin{equation}
\Etrans(x,0)={e^{-c_0x}-e^{-c_0L}\over1-e^{-c_0L}}-{L-x\over L}\;.
\end{equation}
Performing a Fourier analysis and comparing to eq.~(\ref{Etrans}), we
find 
\begin{equation}
a_n=
  {2c_0^2L^2(\cos n\pi-e^{c_0L})\over(-1+e^{c_0L})n\pi(c_0^2L^2+n^2\pi^2)}\;.
\end{equation}
Finally, from this and eqs.~(\ref{conc}) and (\ref{Etrans}) we
conclude that
\begin{equation}
\label{cL}
c_L(t)={2\over L}\sum_{n=1}^{\infty}
 {e^{c_0L}-(-1)^n\over(e^{c_0L}-1)(1+n^2\pi^2/c_0^2L^2)}
  \,e^{-{2n^2\pi^2D\over L^2}t}\;.
\end{equation}
The residual concentration $1/L$ has been eliminated, so that $c_L(t)$
represents the approach to the asymptotic limit (in a single segment of
length $L$, and given that there is at least one particle at the start).  

It is
easily  verified that the limit
$L\to\infty$ yields the known result for the infinite line
\cite{benavraham90}:
\begin{equation}
\label{c_infty}
c_{\infty}(t)=c_0e^{t/t_0}\erfc\sqrt{t/t_0}\;,\qquad t_0={1\over2c_0^2D}\;.
\end{equation}
If $L$ is finite, however, at long times, $t\gg t_1=L^2/(2\pi^2D$, the sum
in~(\ref{cL}) is dominated by the slowest decaying exponential:
\begin{equation}
\label{cL1}
c_L\sim a(L)e^{-t/t_1},\quad a(L)={2\over L(1+\pi^2/c_0^2L^2)}
  {e^{c_0L}+1\over e^{c_0L}-1}\;.
\end{equation}

\section{Systems consisting of segments of random length}
We now have all the tools for the analysis of diffusion-limited coalescence in
a system consisting of a collection of segments of various
lengths, taken from the distribution $\psi(L)$.
The approach to the steady state is given by
\begin{equation}
\label{c(t)}
c(t)=\av{L}^{-1}\int_0^{\infty}c_L(t)\,L\psi(L)\,dL\;,
\end{equation}
where $\av{L}=\int_0^{\infty}L\psi(L)\,dL$ is the average length of the
segments.  Here and henceforth we write $c(t)$ instead of $c(t)-c(\infty)$,
for the sake of brevity.  The residual concentration of particles at the
end of the process, $c(\infty)$, is analyzed in Appendix~A.

Eq.~(\ref{c(t)}), in conjunction with eq.~(\ref{cL}), gives the exact
concentration decay, however, some approximations are necessary in order to
reveal the asymptotic trends.  Thus, we replace the exact result for
$c_L(t)$ with the limiting behavior of eqs.~(\ref{c_infty}) and (\ref{cL1}), 
but expressing the crossover between the two regimes in terms of the length
$\Lambda=\sqrt{2\pi^2Dt}$ rather than the time $t_1$:
\begin{equation}
c_L(t)\sim\cases{
c_\infty(t) &\mbox{$L\gg\Lambda$},\cr
a(L)e^{-t/t_1} & \mbox{$L\ll\Lambda$}.}
\end{equation}
Then,
\begin{eqnarray}
c(t)&\sim&\av{L}^{-1}\Big\{\int_0^{\Lambda}a(L)e^{-t/t_1}L\psi(L)\,dL
        \nonumber\\
&&\mbox{}+c_{\infty}(t)\int_{\Lambda}^{\infty}L\psi(L)\,dL\Big\}\equiv
c_<(t)+c_>(t)\;,
\end{eqnarray}
where $c_<(t)$ and $c_>(t)$ represent the contribution to $c(t)$ from short
($L<\Lambda$) and long ($L>\Lambda$) segments respectively.  Furthermore, a
reasonable approximation to~(\ref{c_infty}) is:
\begin{equation}
\label{c_infty_approx}
c_{\infty}(t)\sim{1\over1/c_0+\sqrt{2\pi Dt}}\;.
\end{equation}
Essentially, at times $t\gg t_0=(2c_0D)^{-1}$ the initial
concentration has no effect and $c_{\infty}\sim1/\sqrt{2\pi Dt}$.

\subsection{Exponential distribution of segment lengths}
Consider first a completely random distribution of segment lengths,
$\psi(L)=\rho e^{-\rho L}$.  This might be the case in chain compounds
when the concentration of disorder (defects along the polymers which
discourage diffusion and effectively terminate the chains) is $\rho$, and the
average segment length is $\av{L}=1/\rho$.  We also assume that $c_0\gg\rho$,
that is, the initial average number of particles per segment is large.  The
contribution from short segments can then be estimated using a saddle point
approximation:
\begin{equation}
c_<(t)\sim 4\sqrt{\pi\over3}\rho\tau^{1/6}e^{-3\tau^{1/3}}\;,
  \qquad \tau\gsim1\;,
\end{equation}
where $\tau=t/(2/\pi^2D\rho^2)$ is a reduced (dimensionless) time.  $c_<(t)$
is essentially the same as the survival probability in the famous trapping
problem \cite{Donsker}, and for similar reasons: the stretched exponential
decay results from the longer survival of particles in very long segments,
weighed against the relative scarcity of such segments. 

The contribution from long segments can be computed analytically:
\begin{equation}
c_>(t)=c_{\infty}(t)[1+\rho\Lambda(t)]e^{-\rho\Lambda(t)}\;.
\end{equation}
Then, with $c_0\gg\rho$, and using the approximation~(\ref{c_infty_approx}),
we have
\begin{equation}
c_>(t)\sim\sqrt{\pi}\rho(1+{1\over2\sqrt{\tau}})e^{-2\sqrt{\tau}}\;,
  \qquad \tau\gsim1\;.
\end{equation}

The two contributions, $c_<$ and $c_>$, become equal at $\tau\approx3.5$. 
Thus, a crossover is expected between the stretched exponential decay with
exponent $1/2$ at early times (but greater than $2/\pi^2D\rho^2$) and a
stretched exponential decay with exponent $1/3$ at later times.  In reality,
the concentration decay is stretched-exponential, with a slowly-varying
exponent; from roughly $1$ at $\tau\approx1$, to $1/2$ at $\tau\approx10$, to
$1/3$ at $\tau\gsim100$  (figure~1).  At the very beginning of the process,
$t\ll2/\pi^2D\rho^2$ ($\tau\ll1$), the survival probability is practically the
same as in an infinite system: $c(t)\approx c_{\infty}(t)$ of
eq.~(\ref{c_infty}), or~(\ref{c_infty_approx}).

\begin{figure}
\centerline{\epsfxsize=8cm\epsfbox{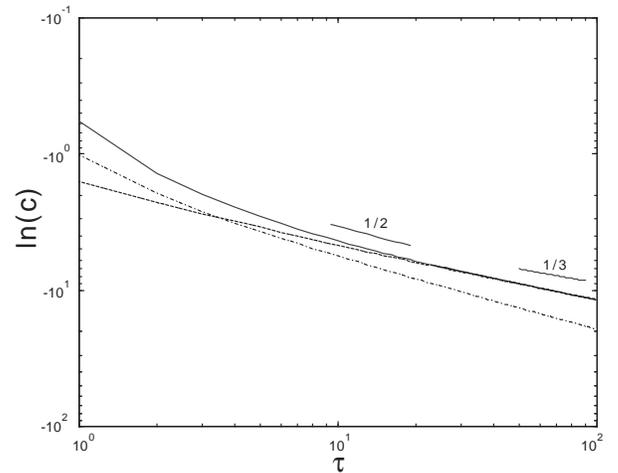}}
\caption{Concentration decay in an exponential distribution of segment
lengths.  Shown is $\ln(c)$ vs.~$\tau$ in a log-log plot, for: $c_<$
(dashes), $c_>$ (dash-dotted line), and $c=c_<+c_>$ (solid line).  Slopes
of $1/2$ and $1/3$ are shown for comparison.}
\end{figure} 

\subsection{Power-law distribution of segment lengths}
Consider next a power-law distribution of the segment lengths:
\begin{equation}
\psi(L)=\cases{
0 & $L<b$,\cr
(\gamma-1)b^{\gamma-1}L^{-\gamma} & $L>b$,  }
\end{equation}
where $b$ is a microscopic cutoff length, and $\gamma>1$.  This is an example
of a long-tailed distribution, and it is also characteristic of percolation
clusters at criticality.   If $\gamma<2$ then the average segment length
diverges (in the thermodynamic limit, when the size of the system is
infinite).  In this case the system is dominated by the largest segment,
which then coincides with the overall size of the system, $L$, and $c(t)\sim
c_L(t)$ of Section~II.  

For $\gamma>2$ the average
segment length is finite: $\av{L}=[(\gamma-1)/(\gamma-2)]b$.
The contribution from short segments can be written as
\begin{equation}
\label{c_small_exp}
c_<(t)=(\gamma-2)b^{\gamma-2}\int_1^{\Lambda/b}a({\Lambda\over u})e^{-u^2}
  ({\Lambda\over u})^{1-\gamma}\,{\Lambda\over u^2}du
\end{equation}
Since $b$ is a microscopic cutoff lengthscale, we assume that $c_0b\lsim1$.
In this case $t\gg t_0=1/(2c_0^2D)$ implies $\Lambda\gg b$.  Due to the term
$e^{-u^2}$, the main contribution to the integral in~(\ref{c_small_exp}) comes
from $u\approx1$, in which case $a(\Lambda/u)\sim2u/\Lambda$, and the upper
integration limit may be extended to $+\infty$.  Thus, 
\begin{equation}
c_<(t)\sim(\gamma-2)\Gamma({\gamma+1\over2},1)
  b^{\gamma-2}(2\pi^2Dt)^{(1-\gamma)/2}\;,
\end{equation}
where $\Gamma(a,x)=\int_x^{\infty}z^{a-1}e^{-z}\,dz$ is the incomplete gamma
function \cite{AS}.

The contribution from the large segments, for $t\gg t_0$ (and $c_0b\lsim1$),
is
\begin{equation}
c_>(t)=c_{\infty}(t)\left({b\over\Lambda}\right)^{\gamma-2}
  \sim  \sqrt{\pi}b^{\gamma-2}(2\pi^2Dt)^{(1-\gamma)/2}\;.
\end{equation}
Thus, the contributions from short and long segments are of the same order of
magnitude at all times $t\gg t_0$, and the overall concentration decays in
power-law fashion, $c(t)\sim(1/b)(b^2/Dt)^{(\gamma-1)/2}$, faster than the
$1/\sqrt{Dt}$-decay of coalescence in the infinite line.
 
\section{Reactions in percolation clusters}
The physical properties of three-dimensional systems with disorder are well
represented by the percolation model of second-order phase transitions
\cite{Sahimi}.  A relevant example is triplet exciton annihilation in
naphtalene-h$_8$ embedded in porous matrices of napthalene-d$_8$
\cite{Kopelman}.  The domains of reaction of naphtalene-h$_8$ can then be
modeled by percolation clusters. Suppose that a reaction domain, {\it i.e.},
a percolation cluster, has mass
$s$.  Then the concentration of particles in the coalescence (or
annihilation) process is \cite{anomalous,Kopelman,Meakin,Hoyuelos}:
\begin{equation}
c_s(t)\sim\cases{
t^{-\ds/2} & $s>\sigma$,\cr
{1\over s}e^{-t/t_1} & $s<\sigma$,}
\qquad t_1\sim s^{2/\ds}, \qquad \sigma\sim t^{\ds/2}\;,
\end{equation}
where $\ds\approx4/3$ is the spectral (or fracton) dimension of the clusters,
and $\sigma$ is the crossover mass between algebraic and exponential decay.

Consider first percolation well below the critical threshold, when the
distribution of cluster masses is $n(s)=\rho e^{-\rho s}$.  The typical
cluster mass $1/\rho$ is smaller the farther one is from the percolation
transition.  Then, following a reasoning similar to the one of Section~III.A,
we find
\begin{equation}
c(t)\sim \rho\tau^{\ds'/2(2+\ds')}e^{-\tau^{\ds'/(2+\ds')}}
  + \rho(1+\tau^{-\ds/2})e^{-\tau^{\ds/2}}\;,
\end{equation}
where $\tau=\rho^{2/\ds}t$ is the reduced time, and the first and second
terms on the r.h.s. represent $c_<$ and $c_>$ respectively.  We note that
the largest (and rarest) clusters in percolation below criticality --- which
are the ones that actually determine the asymptotic behavior of $c_<$ --- are
different from the typical clusters.  They are known as ``lattice
animals", and their spectral dimension $\ds'\approx1.18$ (in $d=3$) is
somewhat smaller than $\ds$.  Thus, the exponent characterizing the stretched
exponential decay of $c(t)$ varies slowly from $\ds/2\approx0.67$ at
$\tau\approx 1$, to $\ds'/(2+\ds')\approx0.37$ at later times.

Exactly at criticality, the distribution of the largest (dominant) clusters is
given by $n(s)\sim s^{-\tau}$, with $\tau$ slightly larger than $2$
($\tau\approx2.186$, in three-dimensional percolation).  Following the
argument for power-law distributions we then find 
\begin{equation}
c(t)\sim t^{-(\tau-1)\ds/2}\;,
\end{equation}
where once again the contributions from small and large clusters are 
comparable.  Using the known values of the various exponents in three
dimensions we predict an algebraic decay: $c(t)\sim t^{-0.79}$.

Well above criticality the system is dominated by the spanning (infinite)
cluster.  This cluster has fractal structure, and is characterized by the
spectral dimension $\ds\approx 4/3$, only for lengthscales shorter than the
correlation length $\xi$.  The correlation length is infinite at criticality,
but shrinks rapidly as one moves beyond the critical threshold.  It is also
the typical size of the {\it finite\/} clusters in the system.  At
lengthscales larger than $\xi$, the spanning cluster is similar to a regular
$d$-dimensional object.  Thus, in three-dimensional percolation above
criticality we find
\begin{equation}
c(t)\sim\cases{
t^{-\ds/2} & $t<\xi^{\dw}$,\cr
t^{-1}     & $t>\xi^{\dw}$,}
\end{equation}
where $\dw\approx3.88$ is the walk dimension of diffusion in the fractal
clusters.  The $1/t$-concentration decay at lengthscales larger than
$\xi$ takes place because the upper critical dimension of diffusion-limited
coalescence (and annihilation) is $d=2$, so that in the effectively
three-dimensional cluster the kinetics is reaction-limited.

\section{Summary and discussion}
Starting from the exact solution of diffusion-limited coalescence (\coal) in
a segment of finite length $L$, we have computed the approach of the
concentration to its steady-state limit, in a system consisting
of disjoint segments characterized by the distribution of their
lengths, $\psi(L)$.  For an exponential distribution, $\psi(L)=\rho e^{-rho
L}$, as might be the case in polymer chains with an element of disorder, we
find three different regimes: (a)~an early time regime where the
concentration decay is similar to that in the infinite line,
$c\sim1/\sqrt{Dt}$, (b) an intermediate time regime, dominated by a
stretched-exponential decay, $c\sim e^{-{\alpha}t^{1/2}}$, and (c)~a
late time regime where $c\sim e^{-{\beta} t^{1/3}}$  ($\alpha$ and $\beta$
are known constants).  In practise, the exponent characterizing the
stretched-exponential decay in regimes~(b) and (c) varies slowly over time,
due to crossover effects and amplitude corrections (figure~1).  It would be
interesting to compare our theoretical results to the anomalous decay of
photogenerated excitons observed in TMMC chains and MX compounds.  We note
that the theory contains no free parameters beyond the diffusion coefficient
of the excitons,
$D$, and the concentration of disorder (defects), $\rho$.  We have also
considered a long-tailed distribution of segment lengths, $\psi(L)\sim
L^{-\gamma}$ ($\gamma>2$).  In this case the concentration decay is $c\sim
t^{-(\gamma-1)/2}$, faster than in the infinite line.

The above results were extended to systems in higher dimensions, where the
reaction zones could be modeled by percolation theory (e.g., annihilation of
triplet-excitons in  naphtalene-h$_8$ embedded in porous matrices of
napthalene-d$_8$).  The results are similar to those in one
dimension, but depend crucially on whether the system is above, below, or
at the critical threshold of percolation, and
$\ds$ (the spectral dimension of the clusters) affects the
actual value of the various decay exponents.  Below criticality, we find an
early time decay of $c\sim t^{-\ds/2}\sim t^{-0.66}$, followed by a
stretched-exponential decay, $c\sim\exp[({\rm const})t^x]$, with a slowly
varying exponent: from $x=\ds/2\approx0.66$ to $x=\ds'/(2+\ds')\approx0.37$
($\ds'$ is the spectral dimension of lattice animals, and all numerical
values cited pertain to the physical case of three-dimensional percolation)
\cite{remark}.  At criticality there is a faster power-law decay, $c\sim
t^{-(\tau-1)\ds/2}\approx t^{-0.79}$, and above criticality there is a
crossover from $c\sim t^{-\ds/2}\approx t^{-0.66}$ at early times to a
reaction-limited decay, $c\sim t^{-1}$, thereafter.

As a final observation, we note that the scaling of the
concentration decay for diffusion-limited annihilation (\annil) is exactly the
same as that of coalescence, and hence the results presented here, in spite of
being grounded on the exactly solvable case of coalescence in a segment, are
valid also for annihilation.  Likewise, the concentration decay at all times,
and in particular the long-time exponential decay of eq.~(\ref{cL1}), remains
essentially the same when the particles are {\it absorbed\/}
at the segment edges, rather than reflected.  Thus, our conclusions are
pretty independent of the fate of the particles as they hit the boundaries
of the reaction domains.

\acknowledgments
We thank Prof.~Harry L.~Frisch for introducing us to the problem, and we
gratefully acknowledge support from the National Science Foundation
(PHY-9820569).

\appendix
\section{Residual particles}

Consider a finite domain of volume $V$ where coalescence or
annihilation might take place.  At the end of the process, when  no further
reactions are possible, there would be left either one or zero
particles.  For coalescence, \coal, the residual number of particles is 
one, unless the volume had been empty to begin with.  In the case of
annihilation, \annil, the parity of the initial number of particles determines
the residue: zero when the initial number is even, and one when it is odd.
 
Suppose that the initial
distribution of particles is completely random, at concentration
$c_0$.  The probability that the domain $V$ contains exactly $n$ particles is 
given by the Poisson distribution:
\begin{equation}
p_n={(c_0V)^n\over n!}e^{-c_0V}\;.
\end{equation}
Thus, for coalescence, the probability that one particle is left over is
\begin{equation}
p_{\rm res}^{\rm coal}=1-p_0=1-e^{-c_0V}\;,
\end{equation}
while for annihilation it is
\begin{equation}
p_{\rm res}^{\rm annil}=\sum_{n{\rm\ odd}}p_n
  ={1-e^{-2c_0V}\over2}\;.
\end{equation}

If the system consists of disconnected domains of reaction, characterized by
a distribution $\psi(V)$ of their volumes, and the initial concentration of 
particles is $c_0$, then the concentration of residual particles at the end of
the process is
\begin{equation}
c_{\rm res}=\av{V}^{-1}\int p_{\rm res}(c_0,V)\psi(V)\,dV\;,
\end{equation}
where $\av{V}=\int V\psi(V)\,dV$ is the average volume of the reaction domains.
For example, if the distribution of reaction-domain sizes is exponential,
$\psi(V)=\rho e^{-\rho V}$, we obtain 
$c_{\rm res}^{\rm coal}=c_0\rho/(c_0+\rho)$ and
$c_{\rm res}^{\rm annil}=c_0\rho/(2c_0+\rho)$.
De~Albuquerque and Lyra \cite{Albuquerque} performed a lattice
computation for the residual concentration of particles for annihilation in
two-dimensional percolation, under the assumption that there is initially one
particle per site.




\begin{thebibliography}{99}
\bibitem{remark} Our results are quite different from \cite{Albuquerque}. 
We note that their numerical simulations for two-dimensional percolation were
for $\tau\lsim20$, not long enough to study the long-time asymptotic behavior.
\end{thebibliography}
\end{document}